# Location-Aware Sign-on and Key Exchange using Attribute-Based Encryption and Bluetooth Beacons

Marcos Portnoi    Chien-Chung Shen
Department of Computer and Information Sciences
University of Delaware
Newark, DE, U.S.A.
{mportnoi, cshen}@udel.edu

*Abstract*—This work presents a mobile sign-on scheme, which utilizes Bluetooth Low Energy beacons for location awareness and Attribute-Based Encryption for expressive, broadcast-style key exchange. Bluetooth Low Energy beacons broadcast encrypted messages with encoded access policies. Within range of the beacons, a user with appropriate attributes is able to decrypt the broadcast message and obtain parameters that allow the user to perform a short or simplified login. The effect is a "traveling" sign-on that accompanies the user throughout different locations.

## I. Introduction

The typical sign-on or login procedure involves a user entering a username and password by means of a keyboard. The backend system verifies the username and password (or the hash of the password) against a database, and then grants, or not, access to a system. Large systems, built upon many components that interoperate, may utilize the single sign-on scheme, in which the user signs on once and gain access to multiple systems.

This concept is normally static, since the login procedure, and the system, assumes the user is at a determined location, from which the user signed on to the system. If the user moves to another location and switches the login terminal, then the previous sign-on is lost and a new sign-on must be performed. If the user signs on from a mobile device, then the user can usually remain logged in to the system, provided a network connection is maintained.

Modern systems may include location and device awareness in various capacities. Microsoft's and Google's free services, for example, can detect the device type utilized and change their interface accordingly (e.g., adjust their service pages to the devices' screen sizes). In addition, in particular in their two-way authentication protocols, each device is individually recognized, and the sign-on procedure can be simplified for previously authenticated devices.

We envision a secure sign-on scheme that allows the login to "travel" with the user, allowing the user to utilize a simplified sign-on procedure based on the user's location. In addition of being location-aware, our scheme employs the expressiveness of Ciphertext-Policy Attribute-Based Encryption (CP-ABE [1]) to encode access policies that are built on both location and user attributes. The access policy is effectively broadcast: the backend system does not need to negotiate access attributes with the user before allowing the simplified sign-on procedure. If the user possesses attributes fulfilling the access policy, then the user will be able to acquire parameters to continue with the simplified sign-on procedure.

Bluetooth Low Energy or Bluetooth Smart beacons construct indoor location information and policy broadcast in our scheme [2]. The user will possess a Bluetooth Low Energy device, such as a compatible smartphone ([3], [4]), to receive the beacons. Outdoor location and policy broadcast may rely on a combination of techniques, such as the same Bluetooth Low Energy beacons, GPS, Wi-Fi, Wi-Fi Direct.

## II. Case Scenario

This scenario details an application of Ciphertext-Policy Attribute-Based Encryption (CP-ABE) for secure location awareness and access control. In this scenario, Bluetooth beacons, installed around an office space (Fig. 1), transmit an encrypted message containing a cryptographic nonce. This nonce is encrypted using CP-ABE and encoding the access rules, through the predicate, that is desired for the range of that beacon.

A user with a Bluetooth device, such as the smartphone, captures the encrypted transmission and attempts to decrypt it utilizing the user's ABE private key. If the user has sufficient attributes to fulfill the encoded predicate in the ABE-encrypted message, then the user device will successfully decrypt the message and obtain the nonce.

To be able to perform a simplified, short-login to a computer within range of the Bluetooth beacon, the user's smartphone must transmit a "location sign-on." The location sign-on procedure is built as follows:

- Hash *nonce + user password*.
- Calculate a new nonce, named c-nonce (for client nonce). The c-nonce is generated according to a token authenticator algorithm. In this case, both the user device and the backend know the c-nonce seed and thus can calculate the same c-nonce for the same time period.
- Encrypt *hash(nonce + password)* using c-nonce as symmetric key.

- Encrypt concatenation of *username + c-nonce(hash(nonce + password))* using nonce as symmetric key.
- Transmit *login[nonce[username + c-nonce(hash(nonce+password))]]* through Bluetooth.

The beacon will receive the transmission, and the backend system will attempt to decrypt the message (the location sign-on) first by decrypting using the current nonce, retrieving the username, then (from the username seed in the backend database) calculating and using the c-nonce as a key to decrypt the rest of the message. If the location sign-on is successful, then the backend system may allow a quick-login for the user, and will in addition know about the location of the user.

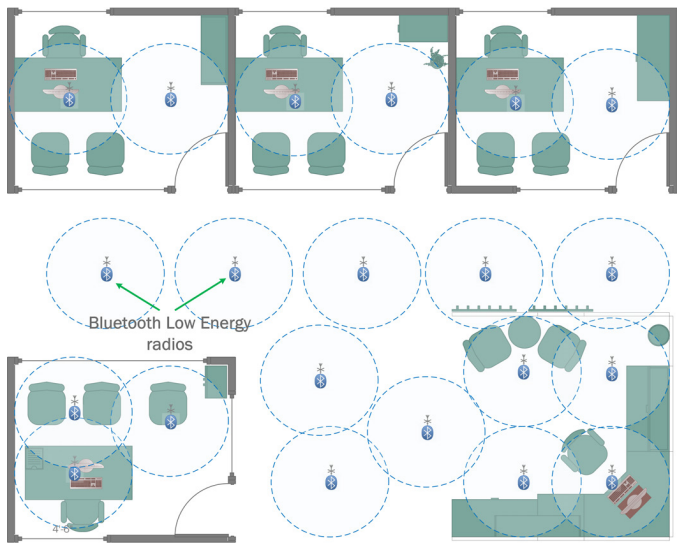

Fig. 1: Bluetooth Low Energy radios distributed around an office.

In the case depicted, the beacon nonce also changes periodically (it can also be generated using a token authenticator algorithm, but the seed is only known by the beacon). The seed for the c-nonce, however, is known by both the user smartphone and the beacon system, and it is unique to a user. The purpose of the beacon nonce is to provide a pseudorandom message that a user device can, or not, decrypt. As it changes periodically, a user device must also submit a new location login periodically, otherwise it will not be possible to perform a quick-login (or not even a full login, depending on the system configuration). The purpose of the c-nonce is to further authenticate the user; since both nonces change periodically, a replay attack is deterred.

It should be noted that the c-nonce is never transmitted from the user device to the beacon. Since the beacon system knows the user seed, it can calculate the current c-nonce and use it as a key to decrypt the user's location login. In addition, when transmitting passwords according to the description here, the passwords are never taken as plaintext, but as hashes of the actual passwords. The only moment at which a password exists as plaintext is when a user is entering the password in a field by means of a keyboard.

### A. The "Location Sign-On" as a Key Exchange Protocol

The "location sign-on" procedure, described earlier, can be summarized as:
1. Broadcast a symmetric key using CP-ABE.
2. Use the symmetric key to encrypt username and an encrypted form of password and nonce.

Effectively, CP-ABE is being used here as a key exchange protocol. We may highlight the following characteristics about this key exchange, and as compared to traditional key exchange formats:

- ABE is expressive, allowing access rules to be encoded in the message itself based on attributes. The message can then be broadcast through insecure medium. The primary access decision, then, need not rely on database access, or extended communication exchange between user and backend, or on specific users (but on "classes" of users). The decision is virtually transferred to the user, as a result of the user being able, or not, to decrypt the broadcast message.
- Access rules can be changed on the fly simply by re-encrypting the new access predicate in the broadcast message.
- The key broadcast using ABE is typically one-to-many, and not one-to-one as traditional key exchanges. It should be noted, however, that some key exchange protocols, such as Diffie-Hellman's, provide key exchange for parties that have no prior knowledge of each other. The method described here requires that ABE private keys be distributed before through secure channels, such that a user can properly decrypt the broadcast messages if this user fulfills the predicate.

### III. PRELIMINARY EXPERIMENTATION AND FUTURE WORK

Initial experimentation focuses on implementing the key exchange algorithm. By means of simulation, we test the effectiveness of the exchange and simplified login scheme when users are within range of different beacon areas. Our future work involves investigating the security of the proposed scheme according to security models, how to address the issues of key revocation and update, and analyze its performance in when subject to several attack vectors.